\documentclass{sigchi}

\toappear{\scriptsize Permission to make digital or hard copies of all or part of this work for personal or classroom use is granted without fee provided that copies are not made or distributed for profit or commercial advantage and that copies bear this notice and the full citation on the first page. Copyrights for components of this work owned by others than the author(s) must be honored. Abstracting with credit is permitted. To copy otherwise, or republish, to post on servers or to redistribute to lists, requires prior specific permission and/or a fee. Request permissions from permissions@acm.org. \\
{\emph{CHI '20, April 25--30, 2020, Honolulu, HI, USA.} } \\
\copyright~2020 Copyright is held by the owner/author(s). Publication rights licensed to ACM. \\
ACM ISBN 978-1-4503-6708-0/20/04\ ...\$15.00.\\
\url{http://dx.doi.org/10.1145/3313831.3376210} }

\usepackage{balance}       
\usepackage{graphics}      
\usepackage[T1]{fontenc}   
\usepackage{txfonts}
\usepackage{mathptmx}
\usepackage[pdflang={en-US},pdftex]{hyperref}
\usepackage[dvipsnames, table]{xcolor}
\usepackage{tabularx}
\usepackage{booktabs}
\usepackage{textcomp}
\usepackage{siunitx}
\usepackage{array}
\usepackage[gen]{eurosym}
\usepackage{cite}

\usepackage{microtype}        
\usepackage{ccicons}          

\usepackage{todonotes}

\usepackage{xspace}

\usepackage{xstring}

\def\plaintitle{Developing a Personality Model for Speech-based Conversational Agents Using the Psycholexical Approach}
\def\plainauthor{Sarah Theres V\"olkel, Ramona Sch\"odel, Daniel Buschek, Clemens Stachl, Verena Winterhalter, Markus B\"uhner, Heinrich Hussmann}

\def\plainkeywords{Big 5; conversational agents; personality.}

\makeatletter
\def\url@leostyle{%
  \@ifundefined{selectfont}{
    \def\UrlFont{\sf}
  }{
    \def\UrlFont{\small\bf\ttfamily}
  }}
\makeatother
\def\pprw{8.5in}
\def\pprh{11in}

\definecolor{linkColor}{RGB}{6,125,233}
\hypersetup{%
  pdftitle={\plaintitle},
  pdfauthor={\plainauthor},
  pdfkeywords={\plainkeywords},
  pdfdisplaydoctitle=true, 
  bookmarksnumbered,
  pdfstartview={FitH},
  colorlinks,
  citecolor=black,
  filecolor=black,
  linkcolor=black,
  urlcolor=linkColor,
  breaklinks=true,
  hypertexnames=false
}

\begin{document}

\title{\plaintitle}

 \numberofauthors{1}
\author{%
  \alignauthor{Sarah Theres V\"olkel$^1$, Ramona Schoedel$^1$, Daniel Buschek$^2$, Clemens Stachl$^3$, Verena~Winterhalter$^1$, Markus B\"uhner$^1$, Heinrich Hussmann$^1$\\
    \affaddr{$^1$LMU Munich, Munich, Germany; $^2$Research Group HCI + AI, Department of Computer Science, University of Bayreuth, Bayreuth, Germany; $^3$Department of Communication, Stanford University, Stanford, US}\\
    \email{sarah.voelkel@ifi.lmu.de}, \email{ramona.schoedel@psy.lmu.de}, \email{daniel.buschek@uni-bayreuth.de}, \email{cstachl@stanford.edu}, \email{verena.winterhalter@campus.lmu.de}, \email{buehner@psy.lmu.de}, \email{hussmann@ifi.lmu.de}}\\
}

\definecolor{rowcol}{rgb}{0.9,0.9,0.9}
\definecolor{Purple}{rgb}{0.64,0.05,0.75}
\definecolor{cr}{rgb}{0.3,0.8,1}
\definecolor{Blue}{rgb}{0,0.2,1}
\definecolor{DanielsColor}{rgb}{0.9,0.6,0.1}
\newcommand{\sarah}[1]{\textsf{\textcolor{Purple}{Sarah: #1}}}
\newcommand{\ramona}[1]{\textsf{\textcolor{Blue}{Ramona: #1}}}
\newcommand{\daniel}[1]{\textsf{\textcolor{DanielsColor}{Daniel: #1}}}
\newcommand{\updatecr}[1]{\textcolor{cr}{#1}}

\newcommand{\PAone}{Confrontational\xspace}
\newcommand{\PAthree}{Dysfunctional\xspace}
\newcommand{\PAfive}{Serviceable\xspace}
\newcommand{\PAeight}{Unstable\xspace}
\newcommand{\PAtwo}{Approachable\xspace}
\newcommand{\PAsix}{Social-Entertaining\xspace}
\newcommand{\PAseven}{Self-Conscious\xspace}
\newcommand{\PAnine}{Social-Assisting\xspace}
\newcommand{\PAfour}{Artificial\xspace}
\newcommand{\PAten}{Social-Inclined\xspace}

\hyphenation{psy-cho-lexi-cal}

\definecolor{topword}{rgb}{0.6,0.6,0.6}
\newcommand{\topword}[2]{#1~{\color{topword}(\expandafter\ifnum #2 < 0 -.\StrRight{#2}{2} \else .#2\fi)}}

\frenchspacing

\maketitle

\begin{abstract}
We present the first systematic analysis of personality dimensions developed specifically to describe the personality of speech-based conversational agents. Following the psycholexical approach from psychology, we first report on a new multi-method approach to collect potentially descriptive adjectives from 1) a free description task in an online survey (228 unique descriptors), 2) an interaction task in the lab (176 unique descriptors), and 3) a text analysis of 30,000 online reviews of conversational agents (Alexa, Google Assistant, Cortana) (383 unique descriptors). We aggregate the results into a set of 349 adjectives, which are then rated by 744 people in an online survey. A factor analysis reveals that the commonly used Big Five model for human personality does not adequately describe agent personality. As an initial step to developing a personality model, we propose alternative dimensions and discuss implications for the design of agent personalities, personality-aware personalisation, and future research.
\end{abstract}




\section{Introduction}

Speech-based conversational agents have become increasingly popular, often presented as helpful assistants in everyday tasks. Due to their intended use and conversational nature, this type of user interface seems much more likely to be seen as a being with a personality, compared to, for example, traditional GUIs~\cite{cafaro2016, krenn2014}. This is further facilitated by recent developments that try to mimic human behavioural characteristics, such as casual filler sounds (``mmm'') on the phone~\cite{Welch2018theverge}.

Such developments highlight that assistants' technical competence alone does not fulfill user needs, for example regarding acceptability in social contexts, yet also considering usability and user experience~\cite{nass2005wired, bouchet2012, andre2000Integrating}: For instance, perceivable personality traits may help to communicate to the users that consistency in the agent's reactions can be expected. 
Recent consumer reports reveal that users particularly enjoy interacting with voice assistants which exhibit a human-like personality~\cite{vlahos2019}. 
More generally, integrating personality into rational agent architectures has been motivated by striving towards more complete cognitive models of agents, as well as sustaining more human-like interactions with people~\cite{bouchet2012}.
These insights motivate the deliberate design of personalities for speech-based conversational agents. 
However, systematically designing agent personalities remains a challenge: For example, conversational agents still struggle with generating adequate human-like ``chitchat'' or humour~\cite{clark2019, porcheron2018}, which often require demonstrations of (consistent) personality.

A crucial missing step towards realising these visions is a scientific model to describe agent personality in the first place. For example, researchers and UX designers could then use such a model to systematically design target personalities for their systems and applications.
Moreover, while basic personality design follows a one size-fits-all approach, future voice assistants can be expected to also aim to adapt to the personality of the individual user. This further heightens the need for a personality model on which such adaptation can be based.

As of today, no dedicated personality model for speech-based conversational agents exists. Thus, most researchers have turned to the Big Five personality taxonomy. Since this model was developed for humans, it remains unclear if it is actually suitable to describe agents. For example, the Big Five dimension of \textit{openness} might be less applicable or important for agents, while new dimensions might be necessary instead (e.g. one capturing an aspect of \textit{artificiality}). 
Moreover, the Big Five taxonomy was developed with a psycholexical approach, that is, using a set of adjectives collected for humans~\cite{deraad2000}. This set might not be adequate for conversational agents.

Recent work supports this view: Zhou et al.~\cite{zhou2019} openly asked about personality of chatbots. Responses included adjectives such as ``robotic'' which are clearly not covered by personality descriptors in models for humans. Similarly, Perez et al.~\cite{perez2019} examined how users perceive brand personality of voice assistants and found out that people attributed ``technical'' or ``logical'' to these assistants.

We address this gap with the first systematic analysis of personality dimensions dedicated to speech-based conversational agents as initial step to developing a personality model. We focus on such agents (also named virtual personal assistants, intelligent assistants, voice user interfaces) due to their prevalent and close interaction with users~\cite{porcheron2018}. 
To find appropriate personality descriptors and dimensions, we apply the psycholexical approach from psychology. 

In particular, we contribute a set of 349 personality descriptors (adjectives), developed with a new multi-method strategy: We combine data from a free description task (online survey, 228 unique descriptors), an interaction task (lab study, 176 unique descriptors), and a text analysis of 30,000 online reviews of Alexa, Google Assistant, and Cortana, which yielded 383 unique descriptors.
As an initial step to developing a personality model for speech-based conversational agents, we contribute ten personality dimensions, derived via exploratory factor analysis on data of 744 people rating our descriptors.
The found dimensions do not correspond to the human Big Five -- neither in number nor content.
We discuss implications for future work and applications of our dimensions and descriptors.

\section{Theoretical Background \& Related Work}
Personality describes relatively stable individual characteristic patterns of acting, feeling, and thinking~\cite{mccrae2008}. Since these patterns are latent traits they cannot be gauged directly. Hence, adequate assessment of peoples' personality has been a major challenge in empirical psychology for a long time~\cite{matthews2003}.

\subsection{Test Construction Process}
The creation of an instrument for assessing psychological constructs (e.g. personality traits) is subject to a complex and multi-stage psychometric construction process \cite{buhner2011, eid2014}. The \textit{Inductive strategy} -- used to develop human personality models~\cite{eid2014} -- assumes no elaborated theory a priori but merely an idea of which items could represent the construct~\cite{buhner2011, eid2014}. An item pool is created, rated by people, and analysed with exploratory factor analysis~\cite{buhner2011}. This statistical method aims to explain associations between items via a small number of homogeneous factors. These factors (i.e. dimensions) can then be used to develop a theoretical model~\cite{buhner2011}. 

A large item pool can be created with different methods~\cite{buhner2011}: The \textit{experience-oriented-intuitive approach} asks people to determine indicators which best describe the construct based on their expertise/intuitive understanding. Another approach is the \textit{collection and analysis of definitions} which reviews the literature for indicators for the construct. The \textit{person-centred-empirical approach} focuses on embedding of the construct in relation to similar constructs. Based on associations of personal characteristics and these constructs in existing literature indicators for the construct of interest are derived. Finally, the \textit{analytical-empirical approach} uses standardised questionnaires, interviews, or observations to identify indicators relevant of the construct \cite{buhner2011}. 
Our multi-method strategy covers multiple such approaches (cf. Our Approach Section).

\subsection{Lexical Approach to Human Personality}
The psycholexical approach~\cite{deraad2000, john1988} is the most established and best known approach in psychology for developing a personality model for humans. It assumes that since people notice and talk about individual differences, these ``will eventually become encoded into their language''~\cite{goldberg1981}.

Allport and Odbert~\cite{allport1936, john1988} collected a large set of 17,953 personality-related terms from a dictionary, Norman~\cite{norman1967} even identified 27,000  unique personality terms. Researchers (e.g.~\cite{allport1936, Cattell1947, goldberg1982, norman1963, norman1967, goldberg1990, john1988}) reduced this by excluding unfamiliar, redundant, etc. terms based on expert and empirical ratings.
In the 1980s Goldberg merged and refined existing lists~\cite{anderson1968, gough1980}. This resulted in 1,710 trait adjectives for self- and peer-report~\cite{john1988}. Goldberg further reduced these to 339 trait adjectives, sorting out synonyms and excluding difficult, slangy, ambiguous, and sex-related adjectives \cite{goldberg1981, goldberg1982, goldberg1990, john1988}.  

A drawback of dictionaries is that language use is not considered (e.g. word frequencies)~\cite{chung2008}. Digital data has enabled such new analyses and researchers have used \textit{self-narrative} texts to extract personality-describing words \cite{chung2008, kulkarni2018}. Our text analysis of online reviews makes use of this idea.

\subsection{The Five-Factor Model of Personality}
The above presented adjective lists have widely been used as basis for exploring the structure of human personality in previous research \cite{john1988}. Despite various approaches to reduce trait terms to few dimensions of human personality across different adjective lists, the \textit{Big Five}, also known as \textit{Five-Factor Model} or \textit{OCEAN}, has emerged as the most prominent model in personality research~\cite{costa1992, mccrae2009, mccrae1985}. 
This model comprises five broad dimensions, which depict individuals' tendencies of feeling, thinking, and behaving~\cite{deraad2000, deyoung2014, diener1992, goldberg1981, jackson2010, jensen2001, matthews2003, mccrae2008, crae1992, mcniel2006}:

\textit{Openness} relates to seeking new experiences, artistic interests, creativity, and intellectual curiosity.
\textit{Conscientiousness} relates to being thorough, organised, neat, reliable, careful, and responsible. 
\textit{Extraversion} relates to being outgoing, sociable, active, and assertive in social interactions. 
\textit{Agreeableness} relates to being friendly, helpful, socially harmonic, kind, trusting, and cooperative. 
\textit{Neuroticism} relates to emotional stability and experiencing anxiety, negative affect, stress, and depression. 

Despite the popularity of the Big Five, personality models are not without limitations. The assessment of personality dimensions via self-report questionnaires is subject to a certain measurement error \cite{gnambs2015}. For example, response biases due to social desirability have been reported \cite{ziegler2009}. Although the Big Five are relatively robust across cultures, there is some controversy whether the model is best suited to describe personality in all cultures. 
\subsection{Describing Agent Personality in HCI}
Humans attribute personality to other humans based on perceptible traces of feelings, thoughts, and behaviour~\cite{scherer1979}. According to the Media Equation~\cite{reeves1996}, people tend to imitate this behaviour for machines and unconsciously assign them personality as well. For example, related work suggests that users make similar personality inferences for speech-based conversational agents as for humans~\cite{mcrorie2012, cafaro2016, krenn2014, isbister2000, neff2010}. More generally, acceptance and credibility of virtual agents are determined by their abilities to be perceived as having a consistent and coherent personality~\cite{mcrorie2012, andre2000Integrating, nass2005wired, tapus2008}. This fundamentally motivates designing agent personality from an HCI perspective.

Today, speech-based conversational agents are most prominently deployed in intelligent personal assistants, such as Google Assistant, Apple's Siri, Microsoft's Cortana, and Amazon's Alexa. Commercially available conversational agents are increasingly used in smart speakers for home automation, e.g. Amazon Echo or Google Home~\cite{porcheron2018}, as well as in automotive user interfaces~\cite{braun2019}. In research, conversational agents have for example been developed to give health advice to users~\cite{devault2014} or serve as social companion for older adults~\cite{Vardoulakis2012}. Braun et al.~\cite{braun2019} showed that users liked and trusted voice assistants more if their personality was matched to the user's own personality. Based on their use context, speech-based conversational agents are likely to show a different personality. For example, a speech-based conversational agent which answers calls at an insurance company might be designed to be reliable and trustworthy. In contrast, a conversational agent in a sports car could be designed as helpful, enthusiastic, and funny. 

McRorie et al.~\cite{mcrorie2012} created speech-based conversational agents based on Eysenck's Three-Factor Model, using behavioural personality cues from the literature. Using a short version of Eysenck's questionnaire, they found that people perceived the agents as intended.
Others used the Big Five model: Several researchers focused on extraversion which has the strongest links to observable behaviour~\cite{krenn2014, isbister2000}. For example, Cafaro et al.~\cite{cafaro2016} evaluated perceived extraversion and friendliness of a virtual museum guide. Moreover, Neff et al.'s findings~\cite{neff2010} indicated that users recognise neuroticism in virtual agents. These projects used questionnaires for human personality to evaluate people's perception of agent personality~\cite{krenn2014, cafaro2012, isbister2000, neff2010}. 
The examples show that perceived agent personality can be deliberately shaped. Yet it is unclear if dimensions of human personality are best suited to inform this since there is no alternative set of dimensions for agents.

Another line of research worked on frameworks for supporting implementation of agent personality: For example, the SOAR (State, Operator And Result) architecture was one of the first attempts to model agent behaviour. The psychological reasoning model BDI (Beliefs, Desires, Intentions) employs personality traits to decide between multiple goals of an agent~\cite{rao1995, rizzo1997personality}. Moreover, Bouchet and Sansonnet~\cite{bouchet2012} developed a computationally-oriented taxonomy for implementing personality traits in voice agents. These frameworks all relied on existing personality models for humans.

In summary, the literature shows example agents with distinctive personality and also frameworks that aim to support their development -- all using \textit{human} personality models. To the best of our knowledge there is no systematic analysis of whether such human models are applicable and sufficiently comprehensive when describing speech-based conversational agent personality. This motivates our investigations in this paper.

There are reasons to expect differences in personality models suitable for humans vs agents: Prior work on human-robot interaction suggests that that there are limitations of the Media Equation since humans conceptualise robots somewhere between alive and lifeless~\cite{bartneck2008exploring}, and the perception of humanness in virtual assistants is a multidimensional construct~\cite{doyle2019}. Furthermore, in an open description of a chatbot's personality, users mentioned descriptors which are not present in the Big Five model, such as \textit{robotic}~\cite{zhou2019}. Since personality is supposed to reflect \textit{distinctive} traits, further dimensions beyond those in human models might be necessary to sufficiently describe agents. This motivates our work on developing a personality model for such agents.

\section{Our Approach}
Similar to the traditional psycholexical approach in psychology, we implemented two steps to derive personality dimensions: (1) \textit{Item Pool Generation}, in which unique phrases for describing personality are collected, and (2) \textit{Exploratory Factor Analysis}, which explores structure and relationship between the items.

\subsection{Item Pool Generation}
The first step finds (English) terms that can ``distinguish the behavio[u]r of one human being from that of another one''~\cite{allport1936}. We seek such terms for agents. 
As adjectives ``are used to describe qualities of objects and persons''~\cite{deraad2000} we limit our set to adjectives, as in prior work~\cite{norman1963, goldberg1982}. We refer to these resulting terms as \textit{descriptors}.  
 
How to best compile a set of descriptors is a key challenge of this approach~\cite{deraad2000}. Inspired by traditional test construction theory~\cite{buhner2011}, we use a new multi-method approach to collect potential descriptors:
\begin{enumerate}
    \item \textit{An online survey}, in which N=135 participants named descriptors for a chosen voice assistant in a free description task \textit{(Experience-oriented-intuitive approach)}.
    \item \textit{A lab experiment}, in which N=30 people interacted with agents (Siri, Alexa, Google Assistant) and described their personality afterwards (\textit{Analytical-empirical approach)}.
    \item \textit{A text analysis} of 30,000 online reviews of agents (Alexa, Google Assistant, Cortana) \textit{(Narrative approach)}.
\end{enumerate}

Since conceptually it is not important how often a descriptor occurs, we collected all unique adjectives regardless of occurrence frequency. We joined the three sets with a given list of descriptors for human personality by Goldberg~\cite{goldberg1990}. We then aggregated the results into a set of 349 adjectives by (1) applying pre-defined exclusion criteria, (2) clustering synonyms, and (3) selecting descriptors based on word-frequency and ambiguity. 

\subsection{Exploratory Factor Analysis}
In this second step, N=744 people rated one of the three most popular assistants (Alexa, Google Assistant, Siri) on the resulting 349 adjectives in an online survey. Exploratory factor analysis on these ratings then revealed latent personality dimensions. 

\section{Descriptors 1: Online Survey}

\subsection{Research Design}
We conducted an online survey to establish a first collection of descriptors for speech-based conversational voice agents. The survey allowed us to get an initial overview to inform subsequent steps; it thus comprised three parts: First, people were asked to indicate with which agents they had interacted before (Siri, Alexa, Google Assistant, Cortana, Samsung Bixby, potential others). Second, participants were asked to provide five representative adjectives to describe the personality of one speech-based conversational agent. Third, participants provided demographic information. 

\subsection{Data Analysis}
We corrected typos (e.g. \textit{helfpul} to \textit{helpful}), replaced nouns with adjective versions where possible (e.g. \textit{fun} to \textit{funny}), and simplified multiple word expressions (e.g. \textit{sometimes annoying} to \textit{annoying}). Furthermore, we excluded all answers which referred to an evaluation of outward appearance (e.g. \textit{good looking}) or usage (e.g. \textit{I don't use it}) instead of personality, as well as further unrelated phrases (e.g. \textit{home button}). 

\subsection{Participants}
We recruited participants via university mailing lists, social media, and online survey communities: 135 participants completed the survey (71.1\% female; mean age 26.2 years, range 18-68 years). 31.1\% of participants had interacted with Alexa before, 69.7\% with Siri, 32.3\% with Cortana, 44.4\% with Google Assistant, and 5.2\% with Samsung Bixby. Only one participant mentioned additional experience with another speech-based conversational agent (\textit{BMW Car Assistant}).

\begin{table}[]
\scriptsize
\begin{tabularx}{\columnwidth}{Xr|Xr}
\toprule
\multicolumn{2}{l}{\textbf{Online survey}} & \multicolumn{2}{l}{\textbf{Interaction experiment}} \\
\midrule
helpful & 34\% & helpful & 90\% \\
friendly & 24\% & friendly & 67\% \\
funny & 19\% & pleasant & 53\% \\
polite & 13\% & funny & 37\% \\
nice & 13\% & likeable & 37\% \\
annoying & 9\% & nice & 37\% \\
calm & 9\% & jolly & 33\% \\
cold & 7\% & polite & 33\% \\
intelligent & 7\% & unpleasant & 33\% \\
fast & 6\% & human & 30\% \\
\bottomrule
\end{tabularx}
\caption{Top ten descriptors mentioned by participants to describe speech-based conversational agents in the online survey (left, N=135) and the interaction experiment (right, N=30). Percentages refer to number of people in the respective study who named each descriptor.}
\label{tab:descriptors}
\end{table}

\subsection{Results}
Our analysis yielded 228 unique descriptors (adjectives): 68.7\% of these were mentioned once. Only five were stated by more than 10\% of people (cf. Table~\ref{tab:descriptors}). In contrast to traditional descriptors for human personality, this set also included adjectives such as \textit{robotic}, \textit{(in)human}, or \textit{impersonal}. 

\section{Descriptors 2: Interaction Experiment}
\subsection{Research Design}
As another approach to collecting descriptors of speech-based conversational agents, we conducted a lab experiment with an interaction task: Here, our goal was to elicit descriptions directly after participants had interacted with such agents.
In particular, we followed a within-groups design, where each participant interacted with three voice assistants (Siri, Alexa, and Google Assistant). 

For each assistant, we asked participants to perform seven tasks, for example to send a message, play a song, or tell a joke. These tasks were informed by related work which analysed the most popular requests to voice assistants at home~\cite{kinsella2018}. We counterbalanced the order of assistants and interaction tasks.  

Interviews followed: To allow participants to become familiar with describing personality characteristics, we first asked them to describe the personality of a friend or family member. Then, after each interaction, participants described the personality of the respective speech-based conversational agent. At the end, they filled in a short questionnaire for demographic data. The experiment took between 45 and 60 minutes. 

\subsection{Data Analysis}
We transcribed the interviews and collected all descriptors from the transcripts. Since the interviews were conducted in participants' native language, two authors individually translated all descriptors, then cross-checked the translations. For standardisation, we turned negations into the corresponding antonyms (e.g. \textit{not funny} to \textit{unfunny}). 

\subsection{Participants}
The sample consisted of N=30 participants (73.3\% female; mean age 24.5 years, range 18-39 years), which were recruited via university mailing lists and personal contacts. 93.3\% of participants knew Siri, 90\% Alexa, 90\% Google Assistant, and 36.7\% Cortana respectively before the study. 87\% of participants interacted at least once with a voice assistant before, and 50\% more than once a week.


\subsection{Results}
The experiment resulted in 176 unique descriptors. Of those 176 descriptors, 110 descriptors were new compared to study one. Again, a majority of descriptors (50.9\%) was only mentioned once. The top ten descriptors can  be found in Table~\ref{tab:descriptors}. Some of the most frequently used descriptors overlap with the ones from the online survey. However, it is interesting to note that more positive affective descriptors were included in the top ten list, such as \textit{likable} and \textit{jolly}. 
\section{Descriptors 3: Text Mining on Online Reviews}

\subsection{Data Acquisition}
In the first two studies, we collected descriptors by explicitly enquiring people about personality traits. In this third study we examined user reviews for descriptors, in order to also include implicit depictions of personality in everyday language use and to cover a wider sample. 
Reviews provide an interesting source as they reflect users' (emotional) experience with an application~\cite{eiband2019, guzman2014, maalej2016}. Following related work~\cite{eiband2019}, we built a web crawler to scrape the latest 10,000 US Google Play Store reviews for Google Assistant, Alexa, and Cortana respectively. We did not include Siri since it is not available in such a store. 

\subsection{Data Processing and Analysis}
Inspired by related work~\cite{eiband2019} on users' problems with intelligent everyday applications, we combined automatic and manual analysis: We first used the Python Natural Language Toolkit\footnote{\url{www.nltk.org}} (NLTK) to automatically extract all adjectives and adverbs from reviews. This resulted in 794 terms for Google Assistant, 913 terms for Cortana, and 1,068 terms for Alexa (incl. intersections).

Adjectives in reviews might not only reflect users' evaluation of personality but also refer to specific features (e.g. \textit{``sucky recognition technology''} or to describe the user (e.g. \textit{``this problem makes me angry''}. Therefore, two authors manually examined all adjectives by going through a random set of reviews per adjective to decide whether this adjective qualified as a descriptor. We excluded descriptors which refer to a state rather than a stable trait (e.g. \textit{``offline''}). For the majority of adjectives, up to twenty reviews were sufficient to decide on exclusion. 
Overall, we included adjectives favourably since our aim was to generate a comprehensive pool of descriptors. Thus, if it was not clear whether an adjective described the speech-based conversational agent or a specific feature, we included it (e.g. \textit{``very useful''}).  

With these criteria, two authors independently went through a random set of hundred adjectives for Google Assistant (corresponds to 886 reviews). The interrater agreement was Cohen's $\kappa = 0.82$. We then compared the results and discussed discrepancies until consensus was reached. The remaining 694 adjectives were split evenly among raters. We repeated this for Alexa and Cortana: Here, we only analysed adjectives which had not been already included for Google Assistant. Again, to ensure interrater consistency, two authors both rated the first 50 adjectives for Alexa and Cortana, respectively. The interrater agreement was $\kappa =0.92$ for Alexa and $\kappa =0.91$ for Cortana. Afterwards, reviews were again split among raters. 

\subsection{Results}
Our analysis yielded 383 unique descriptors; 288 of those were not included in the set from studies one and two. Examples include \textit{busy, clunky, inoperable, laggy, magical, philosophical, romantic, temperamental, usable, virtual}. Given that many reviews are short or only repeat the star rating~\cite{maalej2016, eiband2019}, the number of adjectives seems adequate despite the high number of analysed reviews. Furthermore, we noted that few adjectives occur very frequently (e.g. \textit{helpful} was included in 236 Cortana reviews), while the majority of adjectives appears only occasionally. Since an adjective can occur $n$ times over all 30,000 reviews but may only be used $i\leq n$ times as a descriptor for personality, we deem it not meaningful to show a frequency distribution here.

\section{Final Set of Descriptors}
\begin{figure}[!t]
    \centering
    \includegraphics[width=1\columnwidth]{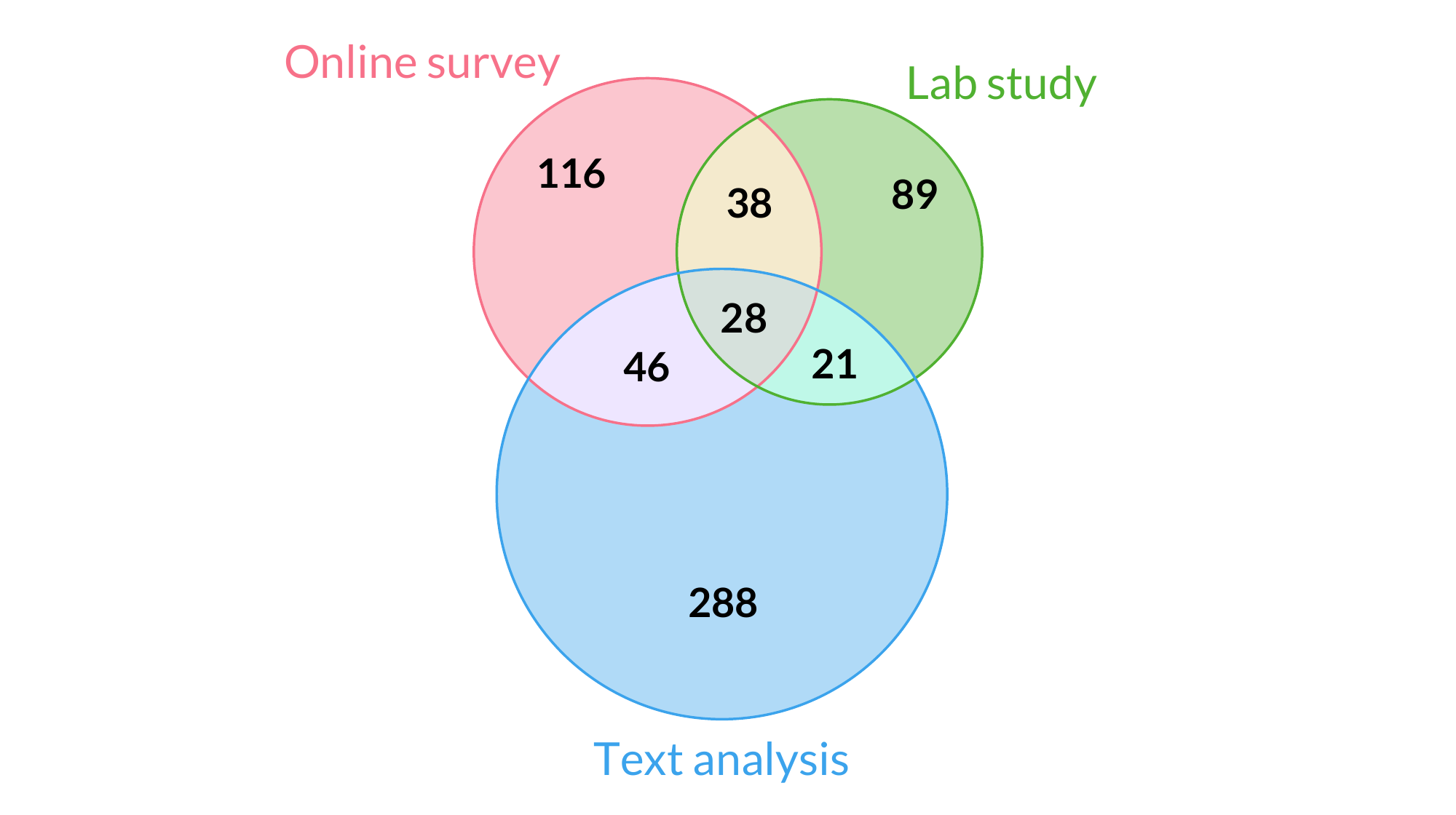}
    \vspace{-1.5em}
    \caption{Overview of the candidate descriptors \textit{before} refinement, as collected with each method and with multiple such methods. The figure shows the value of our multi-method approach: Each of the three methods added new descriptors not found by the others.}
    \label{fig:venn_all}
\end{figure}

As a visual overview, Figure~\ref{fig:venn_all} shows our three obtained sets of candidates (i.e. descriptors before final selection). It is striking that the descriptors collected in the three different methods show only small overlaps. While 28 descriptors were named in all three methods, 493 descriptors were only found in one of the three approaches. We will discuss the implications of this small overlap in the Discussion section. Next we describe how we derived the final set.

\subsection{Adding Existing Descriptors for Human Personality}
We also accounted for the possibility that traditional human personality descriptors may be suitable to describe speech-based conversational agents. Hence, we merged our set with Goldberg's established list of 339 adjectives for human personality~\cite{goldberg1990}. We chose Goldberg's list instead of using the items of personality questionnaires such as the Revised NEO Personality Inventory (NEO-PI-R)~\cite{costa1992} or the Big Five Structure Inventory (BFSI) \cite{arendasy2009bfsi}  since Goldberg's items are openly published so that other researchers can build their work on our item list. The resulting list comprised 870 descriptors. 

\subsection{Refining the Set by Established Exclusion Criteria}
We refined this list in several iterations. In line with the construction of traditional personality inventories in psychology~\cite{goldberg1982, norman1963}, we applied the following exclusion criteria:
slanginess (e.g. \textit{hot}, \textit{screwy}, \textit{snotty});
difficulty (e.g. \textit{antagonistic}, \textit{opportunistic}, \textit{phlegmatic});
ambiguity (e.g. \textit{cool}, \textit{snappy});
link to sex, gender, demographics (e.g. \textit{well educated}, \textit{feminine}, \textit{genderless});
over-evaluation (e.g. \textit{awesome}, \textit{sucky}, \textit{crappy});
peripheral terms (e.g. \textit{dry-witted}, \textit{pseudo-friendly});
anatomical or physical characteristics (e.g. \textit{bulky}, \textit{small}, \textit{beautiful}).


In addition, we also excluded all expressions which refer to the impression on the user rather than the agent's personality (e.g. we include \textit{bored} but exclude \textit{boring}). Finally, in case of lexical opposites (e.g. \textit{dishonest} and \textit{honest}), we only include the positive form since negations have been shown to easily be misunderstood \cite{eid2014}. All exclusion choices were discussed by two researchers and only applied in case of agreement.

\subsection{Refining the Set via Synonym Clusters}
The previous steps resulted in 592 descriptors. Since even comprehensive personality questionnaires usually comprise no more than 300 items for practical reasons~\cite{costa1992, arendasy2009bfsi}, we further reduced the set by removing synonyms. 
This is an established step in related work: Goldberg~\cite{goldberg1982} and Norman~\cite{norman1963} both used expert ratings to exclude redundant terms.
In our case, we instead used a combination of automatic synonym clustering and manual analysis, as described next.

\subsubsection{Specifying Synonyms}
Clustering first required us to specify a list of synonyms for each descriptor. At first, we tried to do so using the lexical database \textit{Word Net}~\cite{miller1995wordnet}. However, the resulting list of synonyms for each word comprised several meanings of that word such that many synonym clusters were not meaningful (e.g. \textit{practical, pragmatic, virtual}).

Therefore, we turned to the online Merriam Webster thesaurus\footnote{\url{www.merriam-webster.com}}, which provides word definitions and synonyms separately for all meanings of a word. We scraped this information. Two authors then manually went through all definitions to compile a list that only included those definitions of a word which are meaningful in the context of personality description. For example, for the descriptor \textit{cold} we included the synonyms for the definition \textit{``having or showing a lack of friendliness or interest in others''} but not for \textit{``having a low or subnormal temperature''}\footnote{\url{www.merriam-webster.com/thesaurus/cold}}. In case a word had $n>1$ valid definitions in this context, we added it $n$ times, with indices ($1...n$). This allowed us to distinguish between the definitions after the clustering. 

\begin{figure}[!t]
    \centering
    \includegraphics[width=1\columnwidth]{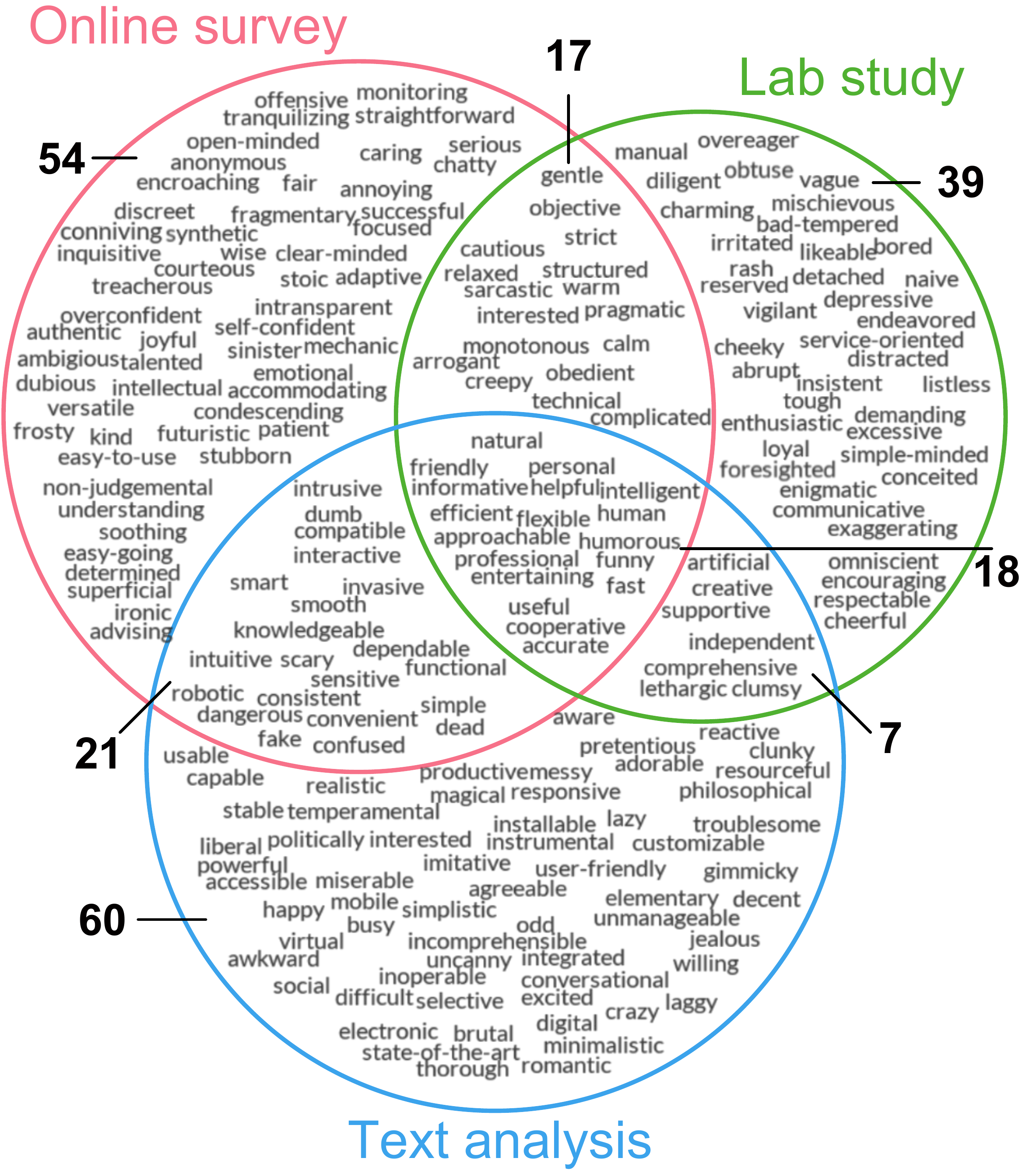}
    \caption{Overview of our final set of 349 descriptors, as collected with each method.} 
    \label{fig:venn_words}
\end{figure}

\subsubsection{Clustering by Synonyms}
We sorted the descriptors by word frequency in the English language with \textit{wordfreq}~\cite{speer2018}. This allowed us to favour frequently used and thus well-known descriptors over unfamiliar expressions.
We used a greedy algorithm that clusters words based on mutual synonymy as follows:
Iterating over the sorted list, at each word $w$, a new cluster $c_w$ is created containing $w$. From the list of synonyms of $w$, all those words $w'$ are added to $c_w$ whose synonym lists also contain $w$ (i.e. mutual synonymy). Finally, $w'$ is removed from the sorted list.


\subsubsection{Final Selection}
We found 230 single descriptors (i.e. clusters of size 1) and 175 synonym clusters (size > 1). For each such cluster, we selected the most frequently used word with only one definition in our set. With this approach, we aimed to maximise clarity of the descriptors overall. For example, for the cluster \textit{aggressive, ambitious, assertive, enterprising} we selected \textit{assertive}; while \textit{aggressive} and \textit{ambitious} are used more frequently they also appeared with other meanings in our overall set. 

We conducted a final manual review and discussion to reduce potentially remaining ambiguity and the number of descriptors with low frequency of use (e.g. which had been selected for clusters with overall low frequency). In this way, we arrived at our final set of 349 adjectives (Figure~\ref{fig:venn_words}).

\section{Exploratory Factor Analysis}

\subsection{Research Design}
We conducted an online survey to examine the structure of the comprehensive set of English trait adjectives we collected using the multi-study approach presented in the previous sections. The final descriptor set of 349 adjectives was administered to participants recruited via Amazon MTurk to get a large and diverse sample \cite{behrend2011, buhrmester2011}. 

After giving informed consent, participants were asked to indicate how often they had already interacted with each of the speech-based conversational agents Siri, Alexa, and Google Assistant. We used different agents to achieve a certain generalisability beyond a specific agent and limited ourselves to the three most common ones because they were most likely to be known by a large number of participants~\cite{voicereport}. Inspired by the approach to assess personality by peer ratings \cite{mccrae1987}, participants were asked to rate the speech-based conversational agent for which they had indicated the highest interaction frequency (minimum criterion was at least once). If they had interacted equally frequently with two or three of them, they were asked to select one of them for the following course of the questionnaire. Afterwards, participants indicated the extent to which each of the 349 adjectives presented in random order applied to the respective selected speech-based conversational agent on a four-point Likert scale ranging from ``untypical'' to ``rather untypical'' to ``rather typical'' to ``typical''. We used an even response scale to avoid a ``neither nor'' category because psychometric research has shown that this is often understood differently by participants and thus leads to problems \cite{eid2014}.
Finally, participants provided demographic information. 

On average, participants took 17 minutes for the online survey and were compensated with \$4.10 \cite{hara2018}. To ensure data quality, we selected MTurk workers only if they met certain requirements regarding their previous work results (80\% accepted work performance in previous studies, at least 1,000 approved work performances), if they were located in the US and indicated to speak English at least well. Following related work~\cite{berinsky2014} we additionally used multiple attention checks throughout the survey. We excluded participants' data for our final analysis if they had missed more than 25\% of the attention tests or their survey time was less than 8 minutes, which we set as the minimum time for serious processing of our questionnaire according to our own preliminary tests. 

\subsection{Participants}
N = 744 participants (45.7\% female, 53.5\% male, 0.8\% said other or preferred not to say) completed the survey and fulfilled our attention/time requirements described above. The mean age was 36.7 (range 19-72 years). 28.6\% indicated to have graduated high school or have a diploma, 17.9\% held an associated degree, and 42.2\% had a bachelor's degree. The remaining 11.3\% had lower or higher educational degrees. 26.1\% of the participants rated Siri, 32.9\% rated Alexa, and 41.0\% rated Google Assistant.

\begin{table}[!t]
\centering
\scriptsize
\renewcommand{\tabcolsep}{0pt}
\renewcommand{\arraystretch}{1} 
\newcolumntype{R}{>{\raggedleft\arraybackslash}X}
\begin{tabularx}{\columnwidth}{rRRRRRRRRR}
\toprule
& \textbf{1} & \textbf{2} & \textbf{3} & \textbf{4} & \textbf{5} & \textbf{6} & \textbf{7} & \textbf{8} & \textbf{9} \\
\midrule
\textbf{\PAone (1)} &  &  &  &  &  &  &  &  &   \\
\textbf{\PAthree (2)} & .42 &  &  &  &  &  &  &  &    \\
\textbf{\PAfive (3)} & -.22 & -.21 &  &  &  &  &  &  &    \\
\textbf{\PAeight (4)} & .64 & .28 & -.19 &  &  &  &  &  &    \\
\textbf{\PAtwo (5)} & .08 & -.01 & .36 & .08 &  &  &  &  &   \\
\textbf{\PAsix (6)} & .24 & .21 & .12 & .17 & .41 &  &  &  &    \\
\textbf{\PAten (7)} & -.21 & -.05 & .43 & -.19 & .42 & .29 &  &  &    \\
\textbf{\PAnine (8)} & .27 & .24 & .14 & .22 & .45 & .35 & .22 &  &    \\
\textbf{\PAseven (9)} & .34 & .31 & .11 & .25 & .28 & .42 & .17 & .32 &    \\
\textbf{\PAfour (10)} & .21 & .25 & -.03 & .19 & .01 & -.04 & -.09 & .19 & .01   \\ 
\bottomrule
\end{tabularx}%
\caption{Correlations between the ten factors from our factor analysis.}
\label{tab:factor_correlations}
\vspace{-1em}
\end{table}

\subsection{Data Analysis}
We investigated the descriptor set's underlying structure with an exploratory factor analysis based on the correlation matrix of all 349 descriptors. To determine the appropriate number of factors we used the empirical Kaiser criterion which has been found to perform well in research settings such as ours~\cite{braeken2017}. We used the R package \textit{psych}~\cite{psych}~(R version 3.6.1).

\definecolor{rowcol}{rgb}{0.9,0.9,0.9}
\begin{table*}[!t]
\centering
\small
\renewcommand{\tabcolsep}{4pt}
\renewcommand{\arraystretch}{2} 
\newcolumntype{L}{>{\raggedright\arraybackslash}X}
\begin{tabularx}{1\textwidth}{l l L}
\toprule
\multicolumn{2}{l}{\textbf{Factor / Label}} & \textbf{Top 20 descriptors (by factor loadings)} \\ 
\midrule
1 & \PAone & \topword{abusive}{71}, \topword{negligent}{70}, \topword{deceitful}{69}, \topword{cruel}{68}, \topword{distrustful}{67}, \topword{combative}{65}, \topword{offensive}{65}, \topword{incomprehensible}{64}, \topword{stingy}{62}, \topword{messy}{60}, \topword{encroaching}{59}, \topword{scornful}{58}, \topword{dubious}{58}, \topword{irritable}{58}, \topword{manipulative}{58}, \topword{explosive}{56}, \topword{clumsy}{55}, \topword{condescending}{54}, \topword{clunky}{54}, \topword{vindictive}{53} \\

\rowcolor{rowcol} 2 & \PAthree & \topword{reckless}{71}, \topword{lazy}{71}, \topword{irritated}{70}, \topword{fearful}{69}, \topword{bitter}{65}, \topword{moody}{65}, \topword{crazy}{63}, \topword{unmanageable}{62}, \topword{prejudiced}{61}, \topword{treacherous}{61}, \topword{ignorant}{59}, \topword{conceited}{59}, \topword{dead}{58}, \topword{inoperable}{57}, \topword{forgetful}{57}, \topword{bad-tempered}{55}, \topword{impetuous}{55}, \topword{listless}{52}, \topword{stubborn}{51}, \topword{egocentric}{50} \\

3 & \PAfive & \topword{informative}{56}, \topword{functional}{56}, \topword{capable}{56}, \topword{accurate}{55}, \topword{knowledgeable}{55}, \topword{convenient}{54}, \topword{thorough}{52}, \topword{responsive}{52}, \topword{productive}{52}, \topword{consistent}{50}, \topword{useful}{50}, \topword{user-friendly}{50}, \topword{interactive}{50}, \topword{adaptive}{50}, \topword{helpful}{50}, \topword{communicative}{49}, \topword{organized}{49}, \topword{dependable}{48}, \topword{intelligent}{47}, \topword{comprehensive}{47} \\

\rowcolor{rowcol} 4 & \PAeight & \topword{nervous}{56}, \topword{depressive}{54}, \topword{rude}{54}, \topword{bigoted}{52}, \topword{grumpy}{51}, \topword{jealous}{49}, \topword{forceful}{49}, \topword{anxious}{48}, \topword{gruff}{47}, \topword{easy-to-use}{-47}, \topword{frosty}{47}, \topword{fretful}{45}, \topword{tempestuous}{45}, \topword{dangerous}{45}, \topword{sloppy}{45}, \topword{faultfinding}{45}, \topword{dumb}{45}, \topword{troublesome}{43}, \topword{rambunctious}{41}, \topword{temperamental}{41} \\

5 & \PAtwo & \topword{peaceful}{64}, \topword{easy-going}{62}, \topword{gentle}{61}, \topword{relaxed}{61}, \topword{fair}{57}, \topword{clear-minded}{56}, \topword{respectful}{53}, \topword{calm}{53}, \topword{humble}{51}, \topword{respectable}{48}, \topword{casual}{48}, \topword{courteous}{47}, \topword{sincere}{46}, \topword{understanding}{44}, \topword{ethical}{44}, \topword{loyal}{43}, \topword{open-minded}{43}, \topword{principled}{42}, \topword{simplistic}{42}, \topword{determined}{39} \\

\rowcolor{rowcol} 6 & \PAsix & \topword{humorous}{69}, \topword{playful}{57}, \topword{funny}{66}, \topword{joyful}{52}, \topword{charming}{51}, \topword{entertaining}{51}, \topword{cheerful}{50}, \topword{merry}{49}, \topword{happy-go-lucky}{49}, \topword{cheeky}{48}, \topword{expressive}{45}, \topword{chatty}{45}, \topword{affectionate}{45}, \topword{happy}{45}, \topword{excited}{44}, \topword{enthusiastic}{44}, \topword{social}{42}, \topword{adorable}{41}, \topword{warm}{39}, \topword{encouraging}{38} \\

 7 & \PAten & \topword{agreeable}{51}, \topword{willing}{46}, \topword{likeable}{43}, \topword{kind}{42}, \topword{trustful}{41}, \topword{decent}{40}, \topword{modest}{40}, \topword{flexible}{40}, \topword{interested}{38}, \topword{realistic}{38}, \topword{conversational}{38}, \topword{friendly}{37}, \topword{self-disciplined}{37}, \topword{inquisitive}{35}, \topword{patient}{34}, \topword{soothing}{32}, \topword{endeavored}{32} \\ 

\rowcolor{rowcol} 8 & \PAnine & \topword{pragmatic}{61}, \topword{fastidious}{58}, \topword{scrupulous}{57}, \topword{genial}{54}, \topword{diplomatic}{46}, \topword{omniscient}{45}, \topword{vigilant}{44}, \topword{vigorous}{44}, \topword{benevolent}{43}, \topword{dignified}{43}, \topword{amiable}{41}, \topword{stoic}{38}, \topword{conscientious}{36}, \topword{discreet}{36}, \topword{deliberate}{35}, \topword{meticulous}{33}, \topword{lenient}{33}, \topword{zestful}{32}, \topword{foresighted}{32}, \topword{restrained}{31} \\

9 & \PAseven & \topword{independant}{45}, \topword{assertive}{45}, \topword{competitive}{43}, \topword{brave}{42}, \topword{creative}{42}, \topword{deep}{41}, \topword{selective}{41}, \topword{proud}{39}, \topword{excitable}{39}, \topword{artistic}{37}, \topword{ambitious}{37}, \topword{introspective}{36}, \topword{powerful}{35}, \topword{individualistic}{35}, \topword{self-indulgent}{35}, \topword{insistent}{35}, \topword{extravagant}{34}, \topword{crafty}{34}, \topword{contemplative}{34}, \topword{daring}{34} \\

\rowcolor{rowcol} 10 & \PAfour & \topword{synthetic}{50}, \topword{robotic}{49}, \topword{intrusive}{48}, \topword{artificial}{47}, \topword{odd}{45}, \topword{invasive}{44}, \topword{gimmicky}{44}, \topword{abrupt}{43}, \topword{superficial}{43}, \topword{monotonous}{43}, \topword{fake}{42}, \topword{simple-minded}{41}, \topword{mechanic}{41}, \topword{vague}{41}, \topword{passionless}{40}, \topword{digital}{39}, \topword{electronic}{39}, \topword{monitoring}{39}, \topword{annoying}{36}, \topword{detached}{35} \\

\bottomrule
\end{tabularx}%
\caption{Overview of the factors obtained in our exploratory factor analysis, with top 20 descriptors (and their factor loadings). The factor labels suggested here are based on the authors' interpretation of all descriptors per factor with loadings of at least .3.}
\label{tab:factor_analysis_results}
\end{table*}

The empirical Kaiser criterion proposed a ten-factorial solution. We used an obliquely (oblimin) rotated principle axis analysis for factor extraction. The resulting ten factors accounted for 49\% of the variance. Table~\ref{tab:factor_correlations} shows their correlations.
Table~\ref{tab:factor_analysis_results} lists the factors with top 20 descriptors by factor loadings and our interpreted factor labels. The complete factor matrix can be found on the project website (cf. Conclusion Section). Loadings ranged between -0.47 and 0.71 across all factors. Using a loading value of 0.30 and high secondary loadings (difference < 0.20) as criteria, 86 of the 349 items could not (uniquely) be assigned to one of the ten factors. 

We next describe each factor as a personality dimension for speech-based conversational agents. We do not claim that ours is the only possible interpretation; readers are invited to develop their own understanding (e.g. via the top descriptors in Table~\ref{tab:factor_analysis_results}). To foster interpretation and discussion, we also sketch ideas and scenarios in which we would expect an agent to score highly on each dimension.

\subsubsection{\PAone}
This dimension is described by negative terms that put the agent into an \textit{actively negative} stance, such as abusive, combative, offensive, stingy, encroaching, manipulative, explosive, or vindictive. Overall, we thus interpret this dimension as capturing a \textit{confrontational} aspect.

For example, a voice assistant scoring high on this dimension might not always readily agree with the user or perform tasks according to the user's wishes. In addition, its feedback might not strike a friendly tone in such situations.

\subsubsection{\PAthree}
This dimension is also described by negative terms, yet puts the agent into a \textit{passive negative} stance. The words signal confusion and inactivity, such as lazy, irritated, fearful, unmanageable, ignorant, dead, inoperable, or forgetful. This inability to function properly is both present on a more emotional/social level (e.g. fearful, listless, conceited, stubborn, crazy) as well as on a  practical/functional one (e.g. inoperable, unmanageable, forgetful, dead). Since both these levels are meanings of the word \textit{dysfunctional}\footnote{\url{www.merriam-webster.com/dictionary/dysfunctional}} we chose this as a label here.

For example, an agent scoring high on this dimension might not react to user input or not give (enough) feedback. It seems likely that voice agents are perceived as highly dysfunctional if their functionality as an assistant is severely hindered, for instance, by software bugs (e.g. resetting mid conversation) or hardware issues (e.g. broken microphone, loss of power).

\subsubsection{\PAfive}
This dimension is described by positive terms, which mostly relate to cognitive functioning, such as informative, functional, capable, accurate, knowledgeable, or thorough. In addition, this dimension also considers adequate communication and role-fulfillment as an assistant, such as convenient, responsive, useful, user-friendly, interactive, communicative, productive, and helpful. Overall, we thus interpret this dimension as capturing functionality and usability of an assistant, which we summarise as \textit{serviceable}.

For example, an assistant scoring high here is likely to react promptly, provides adequate feedback, and performs tasks in a helpful and reliable way. It seems likely that most creators of voice assistants want to present their product as highly serviceable, for example, in advertisements.

\subsubsection{\PAeight}
This dimension is described by negative terms, which overall signal aspects of \textit{instability}, rather than active confrontation or passive inability to function. For example, this includes terms such as nervous, anxious, or temperamental. Besides a level of emotional/social instability present in these terms, there are others that hint more at unstable functionality in this context, such as sloppy, faultfinding, dangerous, and negative (i.e. absence of) ease-of-use.

An example for scoring high on this dimension might be a prototype that does not always work as users expect, with inconsistency being the main negative aspect.

\subsubsection{\PAtwo}
This encompasses positive terms which cast the assistant as calm and welcoming, with words such as peaceful, easy-going, gentle, relaxed, open-minded and understanding. It also hints at an assistant that treats requests well -- fair, clear-minded, respectful, sincere, ethical, loyal, principled, and determined. In contrast to the positive terms of \textit{\PAfive}, these cover not so much the utility of fulfilling tasks as rather the positive social experience expected in asking the assistant to do so. We thus summarise such an assistant character as \textit{approachable}.

A voice assistant scoring high here is likely to navigate well through conversations and gives appropriate feedback that strikes a socially adequate tone, independent of whether the user's request can be practically fulfilled or not.

\subsubsection{\PAsix}
This dimension captures humour in the light of positive social behaviour and entertainment, with terms such as humorous, playful, funny, joyful, charming, entertaining, cheerful, happy, and encouraging.

Scoring high on this dimension likely means that a voice assistant can communicate in a humorous and entertaining way, or includes dedicated functionality for that (e.g. can tell jokes and stories, play games, etc.).

\subsubsection{\PAten}
In this dimension we find an agent's characteristic of being inclined to assist its users, with positive terms such as agreeable, willing, interested, endeavored, flexible, conversational, inquisitive, and patient. This is accompanied by terms that signal a positive tone when communicating this, such as friendly, kind, decent, modest, and soothing.

For instance, a speech-based conversational agent scoring high on this dimension is overall friendly and might actively signal readiness (e.g. with a hardware light) or ask users if they would like a more detailed response or if they have further requests.

\subsubsection{\PAnine}
This dimension captures social skills and attitudes that can be expected from the role of a skilled assistant: It has terms such as pragmatic, conscientious, diplomatic, vigilant, foresighted, amiable, and discreet. It also includes terms that signal accurate task fulfilment such as meticulous, scrupulous and deliberate. In contrast to other dimensions, these relate more to the attitude with which an assistant executes its tasks, rather than its usable functioning (cf. \textit{\PAfive)} or its welcoming character (cf. \textit{\PAtwo}). 

A voice assistant scoring high here likely handles requests well while otherwise staying in the background. It clearly communicates that its role is to serve the user and might also anticipate users' wishes and adequate (re)actions.

\subsubsection{\PAseven}
This dimension encompasses terms that render a voice assistant as an entity capable of independent thought: For example, these terms include independent, competitive, creative, artistic, deep, proud, ambitious, introspective, and contemplative. While actual artificial self-consciousness might still be in the realm of science-fiction for a long time, a voice assistant might create such an illusion in some contexts. This dimension also contains ``magical'' (albeit not in the top 20), which supports such an interpretation.

Designing a speech-based conversational agent to score high here likely requires implementing convincing responses in conversation about opinions, impressions or feelings: In such more abstract conversations the agent might then have room to show shades of (seemingly) independent and creative thought.

\subsubsection{\PAfour}
In this dimension we find terms that emphasise artificiality or ``thingness'': For instance, this covers words such as synthetic, robotic, artificial, gimmicky, superficial, fake, electronic, and mechanic. Other terms here hint at technological implications of bringing such a ``thing'' into human social contexts, such as intrusive, odd, abrupt, simple-minded, passionless, monitoring, annoying, and detached.

A conversational agent might score high on this dimension if it clearly presents itself as an object -- either by communicating this intentionally (e.g. to avoid overtrust) or via issues that break the illusion of an actual being behind the voice.

\section{Discussion}

\subsection{Reflection on Personality Dimensions}
Our multi-method collection of descriptors resulted in ten personality dimensions for conversational agents, derived via exploratory factor analysis on ratings of our descriptors by 744 people (see Table~\ref{tab:factor_analysis_results}).
Reflecting on these dimensions, it is interesting to note that the majority of dimensions indicates either desirable or non-desirable characteristics. This suggests that the agent's ability to fulfill users' expectations of natural conversations is of crucial importance for users' perception. 

Comparing our ten dimensions to the Big Five model for human personality, we find that particularly adjectives from the dimension agreeableness can be found in several of our dimensions, such as \textit{\PAtwo}, \textit{\PAten}, and \textit{\PAnine}. Since we focused our work on voice \textit{assistants}, agreeableness seems to play a key role distributed over several dimensions. 

The dimension \textit{\PAeight} might be associated with the Big Five dimension neuroticism, which is also called emotional stability. Interestingly, this dimension does not only comprise human characteristics of instability, e.g. \textit{nervous} or \textit{depressive} but also technical characteristics such as \textit{inoperable} or absence of \textit{easy-to-use}. Similarly, the dimension \textit{serviceable} encompasses human characteristics, e.g. \textit{knowledgeable}, \textit{thorough}, which correspond with the Big Five dimension \textit{conscientiousness} -- yet also technical terms such as \textit{useful} or \textit{responsive}.  

This contrast of functional and social descriptors appears to be a pattern which can be found within the majority of dimensions: For example, also the dimension \textit{\PAthree} includes descriptors on a social and emotional level (\textit{reckless, moody, crazy}) combined with others that describe the technical functionality or role of an assistant (e.g., \textit{inoperable, forgetful}). The dichotomy of functional and social has also been observed in previous work on how users describe their everyday conversations with voice assistants~\cite{clark2019, perez2019}.

Looking ahead, a different structure might emerge with a greater variety of speech-based conversational agents: For example, while current voice agents predominantly have roles as assistants (which were the subject of our investigation), future agents might fulfill other roles which in turn may impact on their perceived personality and its description. Moreover, the interaction with conversational agents does not really resemble human conversation at the current technological stage, as is underlined by the findings from the interaction experiment (frequent use of descriptors such as as \textit{inhuman}, \textit{impersonal}, or \textit{unpleasant}). However, it is likely that with technological improvements and a more natural conversation in the near future, users might perceive agent personality differently.

Finally, two dimensions emerged that appear as independent of an assistant role: Both \textit{\PAseven} and \textit{\PAfour} seem to describe a speech-based conversational agent's similarity to humans. On the one hand, self-consciousness represents a dimension which current conversational agents cannot technically fulfil, yet the \textit{impression} of for example \textit{creativity} or \textit{independence} can impact on the perceived agent personality. In contrast, Nass and Brave~\cite{nass2005wired} discussed whether speech-based conversational agents \textit{should} be similar to humans. Hence, an agent could appear as self-conscious in its conversation content (e.g. offering opinions) but still highly  \textit{artificial}, e.g. by using a clearly synthesised voice, to inform the user that s/he is communicating with a machine. Doyle et al.~\cite{doyle2019} emphasised that users conceptualise voice assistants' humanness in a multidimensional way. According to their analysis, low interpersonal connection and poor vocal qualities can result in perceiving an agent as artificial, synthetic or robotic but also several of our other dimensions, e.g. humour or kind of knowledge, contribute to the overall perception of humanness.

\subsection{Reflection on Methodology and Implications for Research}

The descriptor sets collected in our three methods show only small overlaps (see Figure~\ref{fig:venn_words}): Only 18 descriptors in the final set were named in all three methods. On the one hand, this could indicate that more data is necessary to derive a more generalisable and robust set. Hence, other approaches or replications are useful and needed, e.g. to evaluate if descriptors ``satuate''. On the other hand, our results show that a combination of multiple methods is beneficial to cover a comprehensive variety of descriptors in different use cases.

We do not regard our resulting set as a ``solution'' but rather as a starting point for future work: For comparison, many researchers have been involved in the collection process for human personality indicators over decades~\cite{deraad2000, john1988}.
A large proportion of the variance of people's answers (51\%) was not explained by our dimensions. This error variance is comparable to results known from human big five models \cite{gnambs2015} and should be addressed in future work. We would also like to point out that the formalisation of a measurement model for agent personality goes beyond the scope of this work. 
We outline key opportunities for such future work here:

Users' purpose to interact with speech-based conversational agents (task vs social conversation use) may influence perception of personality. Since personality is defined as a stable construct across contexts~\cite{mccrae2008}, we started out with a general case to best reflect this definition of personality. In the lab we presented different use cases (task and social). Overall, by combining different methods and including implicit user data (reviews), we collected data from a variety of use cases. Future work should address specific and further use cases.

We focused on speech-based conversational agents. Future work could investigate whether our descriptors can also adequately describe the personality of other agents (e.g., robots, chatbots). To investigate goodness of fit, future work could conduct a \textit{confirmatory} factor analysis using our descriptors for rating these and other agents.
In addition, 25\% of our descriptors could not be clearly assigned to one of dimensions. These items in particular deserve attention in future work. 

We also observed correlations between our dimensions: For example, \textit{\PAone}, \textit{\PAthree}, and \textit{\PAeight} are considerably correlated. These all seem to describe negative aspects of the conversational agent's personality. Similarly, we found a group of dimensions with positive connotations (e.g., \textit{\PAsix}, \textit{\PAten}, \textit{\PAnine}). This could indicate that not only personality traits per se play an important role in assessing the personality of agents, but also their connotation and functional or experience-based evaluation in the user's view. We encourage future research to investigate these aspects separately and take them into account when modelling. For example, it could be investigated if a hierarchical structure of dimensions can be found.

We used English terms (and in the lab study terms translated from German to English by two researchers).
Future work should investigate other languages and cultural backgrounds, since the approach is language-based and different cultures likely perceive agents' personalities differently~\cite{perez2019}.

\subsection{Implications for Practitioners}

An important next step for HCI practice is to derive solution principles for effectively implementing agent personality. Practitioners usually have specific characteristics in mind when designing agent personality. Our descriptors can be used as a \textit{communication tool} to make these characteristics explicit and to discuss the desired personality of a new agent in a systematic way. This seems particularly interesting for collaborations, when multiple conversation designers write dialogues individually, to facilitate consistency and a mutual understanding of an agent personality~\cite{kim2019}.

Related work on the similarity attraction paradigm~\cite{byrne1961} proposed to adapt agent personality to the user~\cite{andrist2015, krenn2014, braun2019}. Agent personality might also be designed with regard to user groups and application context. The found dimensions support such tasks since they make explicit 1) which aspects of personality \textit{can} be varied (e.g. to achieve a goal such as ``neutral, pragmatic helper''), yet also 2) which ones \textit{have to} be considered as well (e.g. \textit{\PAthree} highlights considering that personality is also present in how an assistant communicates failures).

\section{Conclusion}

We presented the first systematic analysis of personality descriptors and dimensions for speech-based conversational agents, following the established psycholexical approach from psychology.
Our main contribution is a set of 349 agent personality descriptors, grouped into ten personality dimensions, which serve as an initial step to developing a personality model for speech-based conversational agents.

As a broader implication, the revealed dimensions do not match the Big Five model. Our descriptors also include terms not associated with human personality. This systematically consolidates evidence from related work about people describing agent personality differently~\cite{zhou2019}. Our findings thus indicate that the human Big Five model is not directly applicable to speech-based conversational agents. Instead, the found personality dimensions also capture, for example, how artificial, self-conscious, or serviceable the agent appears to its users. 

Practically, the found dimensions and descriptors support research and applications in systematically designing personality of speech-based conversational agents.
Conceptually, we set foundations for future work: As in psychology, personality models should be re-validated in further studies. Future work could also investigate models for other virtual agents (e.g. chatbots). Here, our descriptors, dimensions, and methodology may serve as a useful starting point.


To support such future research and applications, our project website hosts the lists of adjectives from the studies, the final descriptor set, and further material from the factor analysis:

\url{www.medien.ifi.lmu.de/personality-model}

\section{Acknowledgements}
This project is funded by the Bavarian State Ministry of Science and the Arts in the framework of the Centre Digitisation.Bavaria (ZD.B).

\balance{}

\bibliographystyle{SIGCHI-Reference-Format}
\bibliography{bibliography}

\end{document}